\documentclass[aps,prl,epsfig,floatfix,showpacs,twocolumn,amssymb,superscriptaddress,amsmath,notitlepage]{revtex4-1}
\usepackage[dvips]{graphicx}
\usepackage{amssymb}
\usepackage{amsmath}
\usepackage{graphicx,color}

\usepackage{float}

\usepackage{enumerate}
\usepackage{mathtools}
\usepackage{stmaryrd}

\usepackage{epstopdf}

\usepackage{textcomp}
\usepackage{gensymb}

\newcommand{\moy}[1]{\left\langle #1 \right\rangle}
\newcommand{\moys}[1]{\langle #1 \rangle}

\newcommand{\ex}[1]{\mathrm{e}^{#1}}

\newcommand{\dd}[0]{\mathrm{d}}

\newcommand{\ee}[0]{\hat{\boldsymbol{e}}}

\newcommand{\RR}[0]{\boldsymbol{R}}
\newcommand{\xx}[0]{\boldsymbol{x}}
\newcommand{\ff}[0]{\boldsymbol{f}}

\newcommand{\pp}[0]{\boldsymbol{p}}

\newcommand{\kB}[0]{k_{\mathrm{B}}}

\newcommand{\nn}[0]{\hat{\boldsymbol{n}}}

\newcommand{\ep}[0]{\epsilon}

\newcommand{\rotop}[0]{\boldsymbol{\mathcal{R}}}
\newcommand{\uu}[0]{ \hat{\boldsymbol{u}}}
\newcommand{\CC}[0]{ {\boldsymbol{\mathcal{C}}}}

\newcommand{\beginsupplement}{%
        \setcounter{equation}{0}
        \renewcommand{\theequation}{S\arabic{equation}}%
     }

\begin{document}

\title{Diffusion of an Enzyme: the Role of Fluctuation-Induced Hydrodynamic Coupling}
\date{\today}

\author{Pierre Illien}
\affiliation{Rudolf Peierls Centre for Theoretical Physics, University of Oxford, Oxford OX1 3NP, UK}
\affiliation{Department of Chemistry, The Pennsylvania State University, University Park, PA 16802, USA}

\author{Tunrayo Adeleke-Larodo}
\affiliation{Rudolf Peierls Centre for Theoretical Physics, University of Oxford, Oxford OX1 3NP, UK}

\author{Ramin Golestanian}
\affiliation{Rudolf Peierls Centre for Theoretical Physics, University of Oxford, Oxford OX1 3NP, UK}

\begin{abstract}
The effect of conformational fluctuations of modular macromolecules, such as enzymes, on their diffusion properties is addressed using a simple generic model of an asymmetric dumbbell made of two hydrodynamically coupled subunits. It is shown that equilibrium fluctuations can lead to an interplay between the internal and the external degrees of freedom and give rise to negative contributions to the overall diffusion coefficient. Considering that this model enzyme explores a mechanochemical cycle, we show how substrate binding and unbinding affects its internal fluctuations, and how this can result in an enhancement of the overall diffusion coefficient of the molecule. These theoretical predictions are successfully confronted with recent measurements of enzyme diffusion in dilute conditions using fluorescence correlation spectroscopy.
\end{abstract}

\pacs{05.40.-a, 82.39.-k, 47.63.mf}
\maketitle

\emph{Introduction.---} The highly precise and efficient functions performed in a biological cell, such as vesicular transport or DNA synthesis, require the conversion of chemical energy into mechanical work by biomolecules \cite{Alberts2014,Phillips2008,Nelson2008}. To this purpose, enzymes and motor proteins perform cyclic turnovers in which they bind to substrate molecules and catalytically convert them to products while undergoing conformational changes, which  affect their transport and diffusion properties. Therefore, the question of whether a biological molecule is able to produce enough mechanical work to overcome the thermal fluctuations of its environment is central for the understanding of biological self-organization and intracellular transport  \cite{Zoback1993,Astumian1997,Julicher1997,Bustamante2001,Kolomeisky2007,Brangwynne2008a,Banigan2011,Howard2001}. 

In this context, fluorescence correlation spectroscopy (FCS) has proven to be a powerful tool to study the physical properties of macromolecules, such as the folding/unfolding or denaturation dynamics of proteins \cite{Sherman2008,Nettels2008a}. Recently, in vitro studies of dilute solutions of enzyme molecules using FCS have revealed that their diffusion coefficient is enhanced when they are catalytically active \cite{Muddana2010a,Sengupta2013,{Sengupta2014a},Riedel2015}. This  phenomenon  may contribute to the self-organization of  biological processes such as the Krebs cycle \cite{Wu2015}. The experimental observation holds for a wide range of enzymes with very different kinetic and thermodynamic properties. Although it was suggested that diffusion enhancement could be correlated to the exothermicity of the reaction catalyzed by the enzyme or its overall catalytic rate \cite{Mikhailov2015,{Bai2015},{Golestanian2015},Riedel2015}, we have recently shown that the slow and endothermic enzyme aldolase could exhibit a similar behaviour \cite{Illien2017a}. The new observations cannot be theoretically explained within the existing nonequilibrium paradigm, and a completely new approach is required. In this Letter, we use a new paradigm to provide a quantitative description for this phenomenon. 


\begin{figure*}
\begin{center}
\includegraphics[width=2.0\columnwidth]{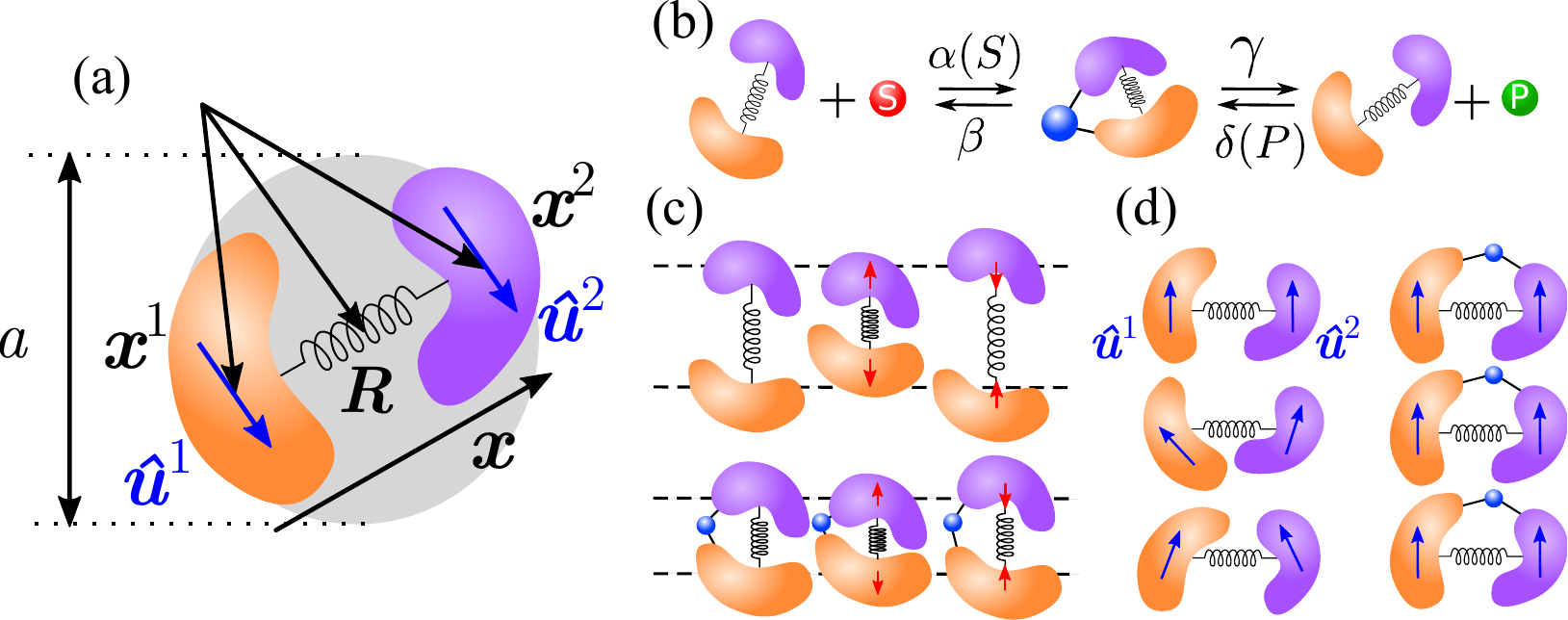}
\caption{
(a) Generalized dumbbell model: two subunits, which represent the modular structure of the enzyme, interact via hydrodynamic interactions and a harmonic-like potential $U$. The conformation of the enzyme is described by the positions of the subunits $\xx^\alpha$ and their orientations $\uu^\alpha$.
(b) Three-state mechanochemical cycle explored by the enzyme in the presence of substrate and product molecules: when it is free, the enzyme can bind to a substrate molecule and transform it into a product molecule which is ultimately released in the bulk. These transformations are assumed to be reversible.
(c) Modification of the extent of elongation fluctuations of the dumbbell due to substrate or product binding. 
(d) Modification of the orientational fluctuations upon substrate or product binding. 
}
\label{conf_changes}
\end{center}

\end{figure*} 

\emph{Main results.---} We propose a simplified description for a generic macromolecular complex using the model of an asymmetric dumbbell, which represents the modular structure of the macromolecule and which reduces its internal degrees of freedom to a minimal number [Fig. \ref{conf_changes}(a)].  Considering the hydrodynamic interactions between the different parts of the enzyme, we show how the internal degrees of freedom affect its overall diffusion. More precisely, we show that the effective diffusion coefficient of the dumbbell has the generic form
\begin{equation}
\label{mainresult}
D_\text{eff} = D_\text{ave} - \delta D_\text{fluc},
\end{equation}
where the first term corresponds to the thermal average of the contributions due to the translational modes of the dumbbell, whereas the second term represents fluctuation-induced corrections arising from the internal elongation and rotation degrees of freedom.  The negative sign of the correction term, which is controlled by the asymmetry of the dumbbell of the individual subunits, is a generic feature of fluctuation-induced interactions \cite{Kardar1999}. 

We then propose a simplified description of the mechanochemical cycle visited by the enzyme [Fig. \ref{conf_changes}(b)], and its influence on the fluctuations of the enzyme. Substrate or product binding generically hampers the fluctuations of the modular structure [Figs. \ref{conf_changes}(c) and \ref{conf_changes}(d)], and therefore reduces the fluctuation-induced contribution to the effective diffusion coefficient. We find that this leads to a relative enhancement of the diffusion ceofficient with a generic dependence on the substrate concentration $S_0$ of the form
\begin{equation}
\label{deltaD}
\frac{\Delta D}{D_0} \equiv \frac{D(S_0)-D(S_0=0)}{D(S_0=0)}=\mathcal{A} \cdot \frac{S_0}{S_0+K},
\end{equation}
where $K$ is the effective equilibrium constant of the chemical cycle, and $\mathcal{A}$ a numerical prefactor that depends on the geometrical and physical properties of the enzyme.

\emph{Model.---} Our goal is to investigate the role played by solvent-mediated hydrodynamic interactions between the different parts that constitute the model enzyme. Real macromolecules generally have a very large number of internal degrees of freedom. In order to describe their inner dynamics, different types of fluctuation modes are to be considered, among which compressional modes, that come from the protein elastic properties, and orientational modes, that originate from the hinge motion of freely rotating subparts. To study these two types of internal degrees of freedom, and in line with previous models of low Reynolds number swimmers \cite{Golestanian2008,Golestanian2010,Sakaue2010}, we reduce the complex geometry of the enzyme to a simple model molecule taking the form a generalized dumbbell made of two subunits, which are geometrically different and whose shapes reproduce the modular structure of the enzyme. They interact via hydrodynamic interactions and via a harmonic-like potential $U$ of stiffness $k$ and equilibrium distance $a$, with a short-distance cutoff that accounts for steric constraints. Their positions $\xx^\alpha$ and orientations $\uu^\alpha$ undergo thermal fluctuations [Fig. \ref{conf_changes}(a)]. This simplified model does not aim to represent a specific enzyme, but will allow us to carry out a detailed analytical study of the fluctuation-induced effects that arise from the hydrodynamic coupling. Although this model can be related to previous attempts to use hydrodynamically-coupled dumbbells or chains to describe the diffusion of polymers \cite{Doi1988}, the question of the internal asymmetries of the dumbbell and of the role played by orientational fluctuations was generally left aside, and our approach offers a novel insight in the fluctuation-induced hydrodynamic coupling between the dumbbell subunits.

\emph{Smoluchowski description.---} The starting point of our analytical treatment is the Smoluchowski equation obeyed by $P(\xx^1,\xx^2,\uu^1,\uu^2;t)$, namely, the probability to find subunit $\alpha$ at position $\xx^\alpha$ and with orientation $\uu^\alpha$ [Fig. \ref{conf_changes}(a)] at time $t$:
\begin{eqnarray}
\label{bigFP}
\partial_t P &=& \sum_{\alpha,\beta=1,2} \left \{  \nabla_\alpha \cdot  \mathbf{M}_\text{TT}^{\alpha\beta} \cdot [ (\nabla_\beta U)P +\kB T \nabla_\beta P ] \right. \nonumber\\
&+& \nabla_\alpha \cdot \mathbf{M}_\text{TR}^{\alpha\beta} \cdot  [  (\rotop^\beta  U) P + \kB T \rotop^\beta P ] \nonumber\\
 &+&  \rotop^\alpha \cdot \mathbf{M}_\text{RT}^{\alpha \beta}  \cdot [(\nabla_\beta U) P + \kB T \nabla_\beta P]  \nonumber \\
  &+& \left. \rotop^\alpha \cdot \mathbf{M}_\text{RR}^{\alpha \beta}  \cdot [(\rotop^\beta U) P + \kB T \rotop^\beta P] \right\},
\end{eqnarray}
where the hydrodynamic tensors $ \mathbf{M}_\text{AB}^{\alpha\beta}$ describe the interactions between translational (T) and rotational (R) modes of the subunits \cite{{Dhont1996},Kim2005}, $U(\xx^1,\xx^2,\uu^1,\uu^2)$ is the interaction potential between the subunits, and the rotational gradient operator is $\rotop^{\alpha} \equiv \uu^\alpha\times \partial_{\uu^\alpha}$ \cite{Doi1988}. We conveniently rewrite Eq. \eqref{bigFP} using the center of mass $\RR = (\xx^1+\xx^2)/2$ and separation (or elongation) $\xx= \xx^2-\xx^1=x\nn$ coordinates. We will use the following type of approximation for any combination of the mobility tensors: $\mathbf{A} \simeq a_0 \mathbf{1}$. This approximation corresponds to the usual pre-averaging of mobility tensors, which consists in neglecting any off-diagonal terms \cite{Zimm1956}, and which has been widely used in polymer physics, and is known to have a wide range of validity \cite{Doi1988}. The effect of the subunits anisotropy, and therefore of the orientation-dependence of the mobility tensors will be the subject of a later publication. The orientation-dependence of the potential $U$ is assumed to take the simple form $U = \frac{1}{2} k (x-a)^2[1  + \sum_{\alpha=1,2} V_\alpha\; \nn\cdot \uu^\alpha + V_{12}\; \uu^1\cdot \uu^2]$, where the dimensionless constants $V_\alpha$ and $V_{12}$ characterize the strength of the constraints on orientational fluctuations, and vanish for freely rotating subunits.

Our aim is to calculate the effective center of mass diffusion coefficient defined as $D_\text{eff}= \lim_{t\to\infty}\frac{1}{6} \frac{\dd}{\dd t} \int_{\RR} \int_{\xx} \int_{\uu^1} \int_{\uu^2} {\RR^2 P}$. A direct attempt to obtain an evolution equation for the second moment $\moy{\RR^2}$ from Eq. \eqref{bigFP} yields an unclosed equation, which includes correlation functions that involve both the external ($\RR$) and internal ($\xx$, $\uu^1$ and $\uu^2$) degrees of freedom. Using Eq. \eqref{bigFP} to write the evolution equation of these quantities, one can show that it involves higher-order correlation functions; therefore, we need an approximation to close this hierarchy of equations. We next identify the hierarchy of the characteristic time scales associated with the different degrees of freedom: from slow to fast we have the center of mass position $\RR$, the orientations $\nn$, $\uu^1$ and $\uu^2$, and the elongation $x$. The elongation relaxation time is indeed $\tau_\text{s} = \xi/k$, where $\xi$ is an estimate of the friction coefficient of the enzyme, whereas the rotational diffusion time of the enzyme $\tau_\text{r}$, such that $\moy{\nn(0)\cdot \nn(t)} \sim \exp(-|t|/\tau_{\text{r}})$, is of the order of $\xi a^2 /\kB T$. Forming the ratio between these two characteristic times yields the dimensionless number $ \tau_\text{s}/\tau_{\text{r}} \sim \kB T/k a^2 \sim \delta x/a$, which is a measure of the relative deformation of the molecule due to thermal fluctuations, and which is smaller than unity.

Guided by this ordering, we average Eq. \eqref{bigFP} over the radial coordinate $x$ assuming that the orientations $\nn$, $\uu^1$ and $\uu^2$ of the dumbbell are fixed. We define the average $\moy{\cdot} = \frac{1}{\mathcal{P}} \int \dd x \, x^2 \cdot P$ where $\mathcal{P} = \int \dd x\,  x^2 P$. The resulting equation satisfied by $\mathcal{P}(\RR,\nn,\uu^1,\uu^2;t)$ is presented in the Supplementary Information \cite{SI}. The next step of the calculation consists of a moment expansion of the equation satisfied by $\mathcal{P}$ with respect to the orientation vectors $\nn$, $\uu^1$ and $\uu^2$, which yields another hierarchy of equations that will need to be approximated by using a closure scheme for orientational order parameters \cite{Saha2014,Golestanian2012,Saintillan2008}. This calculation gives the following leading order expression for the effective  diffusion coefficient \cite{SI}
\begin{eqnarray}
\label{Defflong}
&&D_\text{eff} =  \frac{\kB T}{4}  \moy{m_0} \nonumber\\
&& - \frac{\kB T}{6} \frac{\moy{\gamma_0/x}^2}{\moy{w_0/x^2}} \left[ 1-\sum_{\alpha=1,2} \left( \frac{k a^2}{\kB T} V_\alpha \right)^2 \mathcal{K}_\alpha \right] 
 \end{eqnarray}
where we defined the tensors $\mathbf{M} = \mathbf{M}_\text{TT}^{11}+\mathbf{M}_\text{TT}^{22}+2\mathbf{M}_\text{TT}^{12}$, $\boldsymbol{\Gamma} = \mathbf{M}_\text{TT}^{22}-\mathbf{M}_\text{TT}^{11}$ and $\mathbf{W} = \mathbf{M}_\text{TT}^{11}+\mathbf{M}_\text{TT}^{22}-2\mathbf{M}_\text{TT}^{12}$.
This result has the structure of Eq. \eqref{mainresult}, where the negative fluctuation-induced corrections are  controlled by the coefficient $\gamma_0$, which is a measure of the asymmetry of the dumbbell.  $\mathcal{K}_\alpha$ is a dimensionless coefficient of order 1, that depends on the geometry of the dumbbell and that is estimated to be positive for harmonic-like potentials \cite{SI}.

\emph{Fluctuation-dissipation theorem.---} To validate the above results, we verify that our treatment of the problem satisfies the simplest fluctuation theorem, which takes the form of the usual Einstein relation $\mu_{\rm eff} =  D_{\rm eff}/\kB T$ \cite{Einstein1905a}, where $\mu_{\rm eff}$ is the mobility coefficient of the dumbbell. For simplicity, we do not consider here the effect of the orientation-dependence of the potential $U$. We assume that each of the dumbbell subunits is submitted to an external force $\boldsymbol{f}/2 = (f/2) \ee$ where the amplitude $f$ and the unit vector $\uu$ are arbitrary and constant, so that the total force on the dumbbell is ${f}\ee$. We aim to establish the equation satisfied by the probability distribution of $\RR$ and $\xx$ in the presence of an external force and denoted by $P_{\boldsymbol{f}}(\RR,\xx;t)$. Rewriting Eq. \eqref{bigFP} as $\partial_t P = \mathcal{L}_{\rm T} P$ in order to define the equilibrium Fokker-Planck operator  $\mathcal{L}_{\rm T}$, we find that $P_{\boldsymbol{f}}$ satisfies
\begin{equation}
\label{ }
\partial_t P_{\boldsymbol{f}}= \mathcal{L}_{\rm T} P_{\boldsymbol{f}} - \frac{1}{4} f \nabla_{\RR} \cdot[(\mathbf{M}\cdot  \ee)P_{\boldsymbol{f}}] - f \nabla_{\xx}\cdot [(\boldsymbol{\Gamma} \cdot \ee) P_{\boldsymbol{f}}].
\end{equation}
Following the  averaging procedure presented above, where one can integrate over the radial coordinate $x$ assuming that the orientation is fixed, we first obtain the equation satisfied by $\mathcal{P}_{\boldsymbol{f}} = \int \dd x\, x^2 P_{\boldsymbol{f}}$. Performing again a moment expansion with respect to orientation of the equation satisfied by $\mathcal{P}_{\boldsymbol{f}}$, we get a closed evolution equation of the density $\rho_{\boldsymbol{f}} \equiv \int_{\xx} P_{\boldsymbol{f}}$ as $\partial_t \rho_{\ff} = D_{\rm eff} \nabla^2 \rho_{\ff} - \mu_{\rm eff} \ff\cdot \nabla\rho_{\ff}$, and find that the effective mobility coefficient satisfies the Einstein relation given above.

\emph{Path-integral formulation.---} Given the set of approximations we used to close the hierarchy of equations yielded by the orientation moment expansion, the above calculation only gives the long-time limit of the diffusion coefficient, and does not contain information about  the convergence to this asymptotic result.   In order to get a better insight on the time dynamics of the generalized dumbbell, we turn to a path-integral representation of the stochastic dynamics \cite{Zinn-Justin2002}.  Ignoring for clarity the effect of orientational fluctuations, the starting point is the Langevin equation $\dot{x}^\alpha_i = M_{\text{TT},ij}^{\alpha\beta}F_j^\beta + \sqrt{2 \kB T} \sigma_{ij}^{\alpha\beta}\xi_j^{\beta}$ where the contravariant indices (Greek letters) denote the labels of the particles, and the covariant indices (Roman letters) correspond to the coordinates. The tensors $\boldsymbol{\sigma}^{\alpha\beta}$ are defined as the ``square root'' of the mobility tensors and obey $\sigma_{ik}^{\alpha\gamma} \sigma_{jk}^{\beta\gamma} = M_{\text{TT},ij}^{\alpha\beta}$. The unit white noise $\boldsymbol{\xi}^\alpha$ satisfies $\moy{\xi_i^\alpha(t)}=0$ and $\langle \xi_i^\alpha(t)\xi_j^\beta(t') \rangle =\delta_{ij}\delta^{\alpha\beta}\delta(t-t')$. The force $\boldsymbol{F}^\alpha$ is related to the potential $U$ through $\boldsymbol{F}^\alpha = -\nabla_{\alpha}U$. The propagator conditioned on the starting and arriving points of the dynamics is formally written as the integral of a constraint, and treated following the Martin-Siggia-Rose treatment of such path-integrals \cite{Ros1973}. We obtain an expression in the form $P \propto \int \prod_\alpha \mathcal{D}\xx^\alpha(\tau)\exp\{-\mathcal{S}[\xx^1(\tau),\xx^2(\tau)]\}$, where the action has the simple form
\begin{equation}
\label{ }
\mathcal{S}= \frac{1}{4\kB T} \int \dd\tau \; (\dot{\xx}^\alpha - \mathbf{M}^{\alpha\gamma}_\text{TT} \boldsymbol{F}^\gamma)\cdot \mathbf{Z}^{\alpha\beta}\cdot (\dot{\xx}^\beta - \mathbf{M}^{\beta\delta}_\text{TT} \boldsymbol{F}^\delta).
\end{equation}
Here, the force is $\boldsymbol{F}^\alpha = -\nabla_{\alpha}U$, and the friction tensors $\mathbf{Z}^{\alpha\beta}$ are inverse to the mobility tensors. 

\begin{figure}
\begin{center}
\includegraphics[width=\columnwidth]{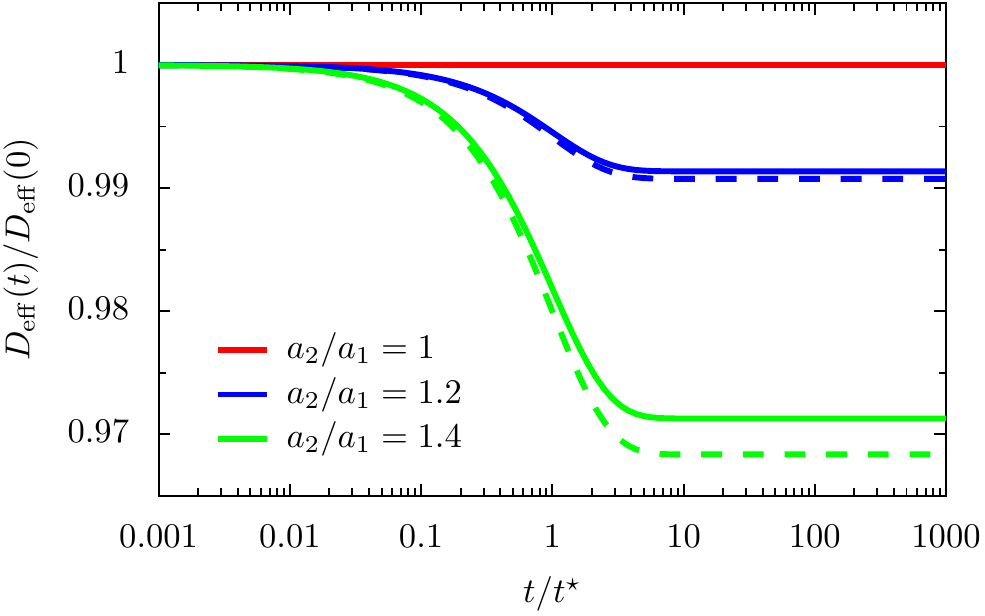}
\caption{Time-dependence of the diffusion coefficient of the dumbbell as obtained by the path-integral formulation. ${D_{\rm eff}}(t)$ is rescaled by its initial value ${D_{\rm eff}}(0)$ and plotted against the rescaled time $t/t^\star$ (see text) for different values of the relative sizes of the protein subunits $a_1$ and $a_2$, for $a_1/a = 0.3$ and $\kB T/(k a^2)=0.01$. For all plots, we compare the results obtained with the mobility functions written in the Oseen limit \cite{Kim2005} (dashed lines) and for spherical subunits with higher-order corrections \cite{Jeffrey1984} (solid lines).}
\label{crossover}
\end{center}
\end{figure}

Using again the diagonal approximation $\mathbf{A} \simeq a_0 \boldsymbol{1}$ as well as the pre-averaging of the hydrodynamic tensors, and changing the variables in the path-integral in order to use the coordinates $\xx$ and $\RR$ instead of $\xx^1$ and $\xx^2$, we obtain the following action:
\begin{equation}
\label{ }
\mathcal{S}  = \frac{1}{\kB T} \int \dd \tau  \frac{\moy{z_0} }{4} \dot{\RR}^2  +\mathcal{S}_{\xx} , 
\end{equation}
with
\begin{equation}
\label{ }
\mathcal{S}_{\xx}  = \frac{1}{\kB T} \int \dd \tau \left[\frac{\moy{y_0} }{16}\dot{\xx}^2 +\frac{\moy{\zeta_0}}{4}  \dot{\xx}\cdot\dot{\RR} + \moy{w_0} U'^2  \right], 
\end{equation}
where we define $\mathbf{Z} = \mathbf{Z}^{11}+\mathbf{Z}^{22}+2\mathbf{Z}^{12}$, $\mathbf{Y} = \mathbf{Z}^{11}+\mathbf{Z}^{22}-2\mathbf{Z}^{12}$, and $\boldsymbol{\zeta} = \mathbf{Z}^{22}-\mathbf{Z}^{11}$. For a simple harmonic potential $U=k \xx^2/2$, the $\xx$-dependent part of the action $\mathcal{S}_{\xx}$ can be written as the time integral of a quadratic form.  After integration over the paths $\{\xx(\tau)\}$, which corresponds to an integration of the fast degrees of freedom of the dumbbell, this yields
\begin{equation}
\label{ }
P \propto \int \mathcal{D} \RR(\tau) \; \exp\left \{ -\frac{1}{4\kB T} \int \frac{\dd \omega}{2\pi}  \frac{\omega^2 |\RR(\omega)|^2}{\mu(\omega)} \right\},
\end{equation}
where we have written the integral over $\tau$ in Fourier space and defined a mobility as $\mu(\omega)^{-1} = \moy{z_0}-(\moy{\zeta_0}^2\omega^2/4)\widetilde{G}(\omega)$, with the Green's function being $\widetilde{G}(\omega) = (\moy{y_0}\omega^2/4+\moy{w_0}k^2)^{-1}$. We deduce the mean square displacement of the center of mass using the Green-Kubo relation \cite{Kubo1957}, and find
\begin{equation}
\label{msd}
\moy{\RR^2} = 6 \; \frac{\kB T}{\moy{z_0}}\left[t+ \frac{\moy{\zeta_0}^2}{\moy{z_0}\moy{y_0}-\moy{\zeta_0}^2}t^\star(1-\ex{-t/t^\star})\right].
\end{equation}
A time-dependent diffusion coefficient, defined as $D_{\rm eff}(t) = \frac{1}{6} \frac{\dd}{\dd t}\moy{\RR^2}$, is deduced, and relating the averaged resistance tensors $\moy{z_0}$, $\moy{y_0}$ and $\moy{\zeta_0}$ to the mobility tensors \cite{SI}, we find the following asymptotic regimes
\begin{equation}
\label{ }
D_{\rm eff} = \begin{cases}
  \frac{\kB T}{4} \moy{m_0}    & \text{;  $t \ll t^\star$ }, \\
 \frac{\kB T}{4} \left[\moy{m_0}-\frac{\moy{\gamma_0}^2}{\moy{w_0}}   \right]  & \text{;  $t \gg t^\star$ },
\end{cases}
\end{equation}
where the crossover time $t^\star= (\moy{w_0}k)^{-1}$ is the elongation relaxation time. The drop in the diffusion coefficient between the two limiting regimes is therefore  $-\delta D_{\rm fluc} = -\frac{\kB T}{4} \frac{\moy{\gamma_0}^2}{\moy{w_0}}$, which is in agreement with the result obtained from the Smoluchowski description of the dynamics [Eq. \eqref{Defflong}], where we showed the existence of a negative correction to the diffusion coefficient of the dumbbell due to its asymmetry. We therefore confirm this observation, and highlight the consistency between the two treatments of the stochastic dynamics. The full time dependence of the effective diffusion coefficient is plotted in Fig. \ref{crossover}.

\emph{Mechanochemical cycle.---} When the enzyme is placed in the presence of substrate molecules, it will go through a mechanochemical cycle: depending on whether the enzyme is free, bound to a substrate molecule or bound to a product molecule, its conformation will fluctuate around different equilibrium states. In other words, substrate binding and unbinding can strongly affect the fluctuations of the internal degrees of freedom and therefore impact the overall diffusion coefficient of the enzyme, as unveiled by our simple model. In order to go further with this idea, we consider the simplified three-state catalytic cycle represented in Fig. \ref{conf_changes}(a): the enzyme binds to a substrate molecule, and converts it into a product molecule that is ultimately released. Both steps are assumed to be reversible.

From a conformational point of view, the enzyme then only exists in two states where it is respectively free or bound, characterized by the potentials $U_\text{f}(\CC)$ and $U_\text{b}(\CC)$, where $\CC$ is a high-dimensional vector describing the conformation of the enzyme. It is straightforward to show that, within this discrete-state equilibrium picture, the average of any conformation-dependent function writes 
$\moy{\Phi} = \moy{\Phi}_\text{f}+f(S,P)\left[\moy{\Phi}_\text{b}-\moy{\Phi}_\text{f}\right]  $, 
where $f$ is a function of the substrate and product concentrations \cite{SI}:
\begin{equation}
\label{ }
f(S,P) = \frac{S}{S+\frac{K_\text{S}}{K_\text{P}}P+K_\text{S}} + \frac{P}{P+\frac{K_\text{P}}{K_\text{S}}S+K_\text{P}}.
\end{equation}
The effective equilibrium constants $K_\text{S}$ and $K_\text{P}$ are defined as
\begin{eqnarray}
\label{ }
K_{\text{S}} &=& K_{\text{S},0}\frac{\int_{\boldsymbol{\mathcal{C}}} \ex{-U_\text{f}/\kB T}}{\int_{\boldsymbol{\mathcal{C}}} \ex{-U_\text{b}/\kB T}},\\
K_{\text{P}} &=& K_{\text{P},0}\frac{\int_{\boldsymbol{\mathcal{C}}} \ex{-U_\text{f}/\kB T}}{\int_{\boldsymbol{\mathcal{C}}} \ex{-U_\text{b}/\kB T}},
\end{eqnarray}
where $K_{\text{S},0}$ and $K_{\text{P},0}$ are the bare equilibrium constants of substrate and product binding respectively. We assume for simplicity that the chemical rates $\alpha(S)$ and $\delta(P)$ defined in Fig. \ref{conf_changes}(b) are linear functions of $S$ and $P$ respectively, with $ \alpha(S) = \alpha_0 S$ and $\delta(P) = \delta_0 P$. The bare equilibrium constants then read $K_{\text{S},0}=\beta/\alpha_0$ and $K_{\text{P},0}=\gamma/\delta_0$ \cite{SI}.

 We must emphasize that this simplified mechanochemical cycle neglects  the nonequilibrium step of the reaction where substrate molecules are actually turned into product molecules. The reversible binding and unbinding steps can indeed be assumed to happen faster than the  chemical step \cite{Rago2015,Illien2017a}. The equilibrium picture we present here is then valid at any stage of the chemical reaction, whether the system is in a transient state or has reached chemical equilibrium. Therefore, the expression of the diffusion enhancement presented in Eq. \eqref{deltaD} is valid both at the early stages of the reaction, where very few substrate molecules have been converted into product and where the effective equilibrium constant is $K=K_\text{S}$, and at chemical equilibrium, where $K=(K_\text{S}+K_\text{P})/2$ \cite{SI}. In the experiments reported in the introduction, FCS measurements are performed on a short timescale right after the enzyme is placed in the presence of its substrate. Comparing this almost stationary measurement of the early stage of the reaction with what could be observed once chemical equilibrium is reached could bring information about the relative affinity of the enzyme with the substrate and product molecules, and about the  magnitude of the corresponding conformational changes. This will be the object of future work.

The diffusion coefficient of the dumbbell is  given by Eq. \eqref{Defflong}, where the averages of the mobility functions are calculated using the expression above. Since the modifications due to substrate binding are expected to be relatively small, we can expand the resulting expression in powers of the relative change in the configuration averages $\left[\moy{\Phi}_\text{b}-\moy{\Phi}_\text{f}\right]/ \moy{\Phi}_\text{f}$ and stop at the lowest order, which will provide us with an expression of the form of Eq. \eqref{deltaD} for the relative change in diffusion coefficient, where the dimensionless quantity $\mathcal{A}$ is a function of various mobility coefficients in the free and bound states, but not the substrate concentration \cite{SI}. The simple expression we obtain for $\Delta D/D_0$ is to be compared with previous experimental measurements. First, the Michaelis-Menten-like dependence on the total substrate concentration $S_0$ corresponds to the experimental observations that we recently reported \cite{Illien2017a}. Secondly, since $\mathcal{A}$ is constructed as a ratio between averages of similar quantities, by default we expect it to be of order unity, which is confirmed by the measured relative diffusion enhancement $\mathcal{A}$ at substrate saturation ($S_0 \gg K$), found to be of the order of a fraction of unity.

\emph{Conformational changes.---} In order to be more specific, we will finally investigate  the consequences of a number of typical structural modifications on the diffusion coefficient of the enzyme. First, taking the example of aldolase \cite{Rago2015}, binding will influence the average conformation  by bringing the two subunits closer by a few \AA. This can be incorporated by choosing $U_\text{f} = \frac{1}{2}k_\text{f}(x-a)^2$, and $U_\text{b} = \frac{1}{2}k_\text{f}[x-(a-\delta x)]^2$, where $a$ is the equilibrium distance between the subunits (of the order of a few nm), and where $\delta x$ represents the typical displacement of the protein residues between different conformational states [Fig. \ref{conf_changes}(b)]. The substrate molecule may also play the role of a stiff cross-linker for the protein, and binding is likely to increase significantly the effective stiffness of the interaction potential, that will read $U_\text{b} = \frac{1}{2}k_\text{b}(x-a)^2$ with $k_\text{b} \gg k_\text{f}$. Finally, substrate binding will also affect the fluctuations of the orientational modes of the dumbbell through the coefficients $V_{\alpha}$. Such an effect was for instance suggested for enzymes like urease, which is known to have a `flap' that is closed when the enzyme is bound and open otherwise \cite{Roberts2012}. 
 
 \begin{figure}
\begin{center}
\includegraphics[width=0.9\columnwidth]{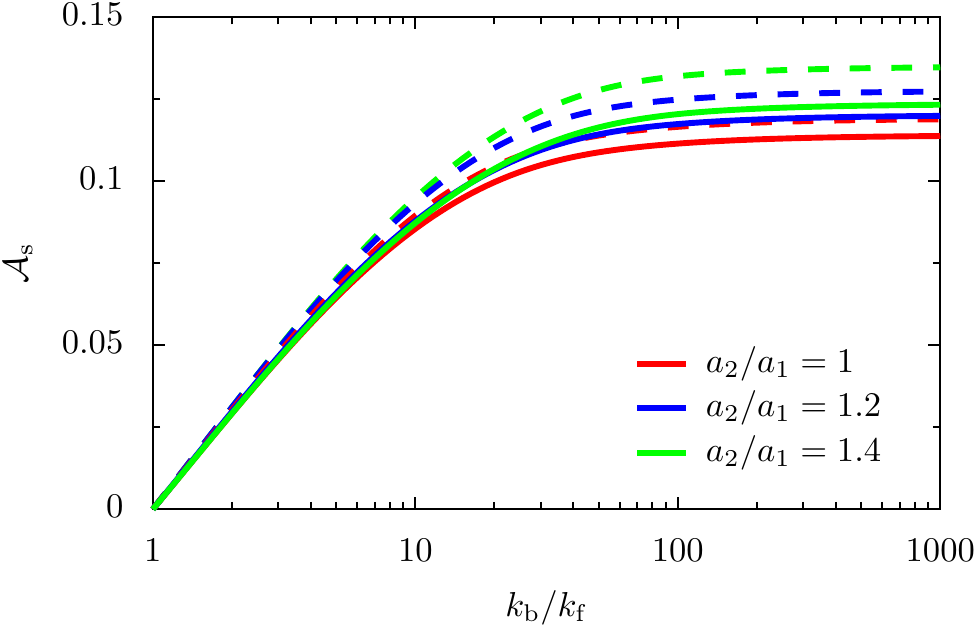}
\caption{The contribution to the amplitude $\mathcal{A}_\text{s}$  when substrate binding modifies the stiffness of the interaction potential from $k_\text{f}$ to $k_\text{b}$. In the free state, the potential stiffness is taken as $k_\text{f}=\kB T/a^2$. The mobility functions are written in the Oseen limit \cite{Kim2005}  (dashed lines) and for spherical subunits of radii $a_1$ and $a_2$ with higher-order corrections \cite{Jeffrey1984} (solid lines). We compare different values of the asymmetry $a_2/a_1$, and the equilibrium length of the potential is such that $a_1/a=0.3$.}
\label{stiffnessfig}
\end{center}
\end{figure}
For concreteness, we now estimate the contribution to $\mathcal{A}$ by the different specific structural modifications of the enzyme discussed above. We first focus on the compressional modes of fluctuations and on the associated terms in Eq. \eqref{Defflong}. When the change in the diffusion coefficient originates from a reduction in the equilibrium distance between the subunits, we can expand the expression for $\mathcal{A}$ in the limit of small relative deformation ($\delta x \ll a$) and large potential stiffness ($k_\text{f} \gg \kB T/a^2$) \cite{SI} to find $\mathcal{A}_\text{c}= \mathcal{G} \cdot {\delta x}/{a}$ with the dimensionless parameter defined as
\begin{equation}
\label{ }
\mathcal{G} = a \left[ -m_0' +\frac{1}{6} \frac{\gamma_0^2}{w_0}  \left( \ln \frac{\gamma_0^2}{w_0} \right)'\right] \left[m_0 - \frac{1}{6}\frac{\gamma_0^2}{w_0}  \right]^{-1},
\end{equation}
where the mobility tensors are evaluated at the equilibrium length $x=a$ (i.e. the configuration averaging has been implemented). While $\mathcal{G}$ is a purely geometric quantity that is typically of order unity, the result shows that the extent of the average deformation directly controls the magnitude of $\mathcal{A}$. For the case when the substrate binding makes the protein stiffer, Fig. \ref{stiffnessfig} shows a plot of the relevant factor $\mathcal{A}_\text{s}$ as a function of the ratio between the two stiffnesses $k_\text{b}/k_\text{f}$; we observe that this factor is typically of order one and that it increases when the two stiffnesses deviate from one another substantially. We can also find a closed form expression in the limit of very large $k_\text{f}$ and $k_\text{b}$ with a finite difference $\delta k$ as $\mathcal{A}_\text{s}= \mathcal{H} \cdot [\kB T/(k_\text{f} a^2)] \cdot ({\delta k}/{k_\text{f}})$ where $\mathcal{H}$ is a dimensionless coefficient of order unity \cite{SI}. Finally, if the orientational fluctuations of the dumbbell are affected by substrate binding, we can deduce the  contribution to the relative change of the diffusion coefficient in the simple case where the subunits are freely fluctuating in the free state ($V_{\alpha}=0$) and constrained otherwise ($V_{\alpha}>0$):
\begin{equation}
\label{ }
\mathcal{A}_\text{r} \simeq \frac{\kB T }{k_\text{f} a^2} \sum_{\alpha=1,2} V_{\alpha,\text{f}} (V_{\alpha,\text{b}}-V_{\alpha,\text{f}}) \mathcal{J}_\alpha,
\end{equation}
where $\mathcal{J}_\alpha$ are dimensionless coefficients of order unity \cite{SI}.

\emph{Conclusion.---} We have proposed a simple model to study the effect of asymmetry on the fluctuation-induced hydrodynamic coupling between the different parts of a model enzyme. We consider the interplay between its internal and external degrees of freedom and calculate the corrections to the overall diffusion coefficient that originate from the compressional and orientational degrees of freedom, and that are controlled by the structural asymmetries of the molecule. This generic model can be used to describe the mechanochemical cycles explored by enzyme molecules when placed in the presence of substrate molecules. We show how substrate binding and unbinding can lead to diffusion enhancement, and confront our theoretical predictions to recently published experimental measurements. Our minimal model, that contains all the required physical ingredients, completes the existing theoretical picture that failed to explain consistently the experimental observations so far.

\acknowledgments
P. I. and R.G. acknowledge financial support from  the US National Science Foundation under MRSEC Grant No. DMR-1420620. We benefited from fruitful discussions with A. Sen and K. K. Dey.


%

\onecolumngrid

\pagebreak

\beginsupplement

\begin{center}

\textbf{Diffusion of an Enzyme: the Role of Fluctuation-Induced Hydrodynamic Coupling}

$\ $

\textit{\textbf{Supplementary Information}}

$\ $

 Pierre Illien,$^{1,2}$ Tunrayo Adeleke-Larodo,$^1$ Ramin Golestanian,$^1$

$\ $

$^1$\emph{Rudolf Peierls Centre for Theoretical Physics, University of Oxford, Oxford OX1 3NP, UK}

 $^2$\emph{Department of Chemistry, The Pennsylvania State University, University Park, PA 16802, USA }

\end{center}

\section{Combination of mobility tensors used in the main text}

\begin{eqnarray}
\label{ }
\mathbf{M}(\uu^1,\uu^2,\nn) & = &    \mathbf{M}_\text{TT}^{11}+\mathbf{M}_\text{TT}^{22}+2\mathbf{M}_\text{TT}^{12}\simeq [m_0 + m_1 (\nn \cdot \uu^1) +m_2( \nn \cdot \uu^1) +m_{12} (\uu^1 \cdot \uu^2)  + \dots] \; \mathbf{1}, \label{exp1}\\
\mathbf{W}(\uu^1,\uu^2,\nn) & = &    \mathbf{M}_\text{TT}^{11}+\mathbf{M}_\text{TT}^{22}-2\mathbf{M}_\text{TT}^{12} \simeq [w_0 + w_1 (\nn \cdot \uu^1) +w_2( \nn \cdot \uu^1) +w_{12} (\uu^1 \cdot \uu^2)  + \dots] \; \mathbf{1}, \\
\boldsymbol{\Gamma}(\uu^1,\uu^2,\nn) & = &   \mathbf{M}_\text{TT}^{22}-\mathbf{M}_\text{TT}^{11} \simeq [\gamma_0 + \gamma_1 (\nn \cdot \uu^1) +\gamma_2( \nn \cdot \uu^1) +\gamma_{12} (\uu^1 \cdot \uu^2)  + \dots] \; \mathbf{1}, \\
\boldsymbol{\Psi}^{(1)}(\uu^1,\uu^2,\nn) & = &    \mathbf{M}_\text{RR}^{11} \simeq [\psi^{(1)}_0 + \psi^{(1)}_1 (\nn \cdot \uu^1) +\psi^{(1)}_2( \nn \cdot \uu^1) +\psi^{(1)}_{12} (\uu^1 \cdot \uu^2)  + \dots] \; \mathbf{1}, \\
\boldsymbol{\Psi}^{(2)}(\uu^1,\uu^2,\nn) & = &    \mathbf{M}_\text{RR}^{22}\simeq [\psi^{(2)}_0 + \psi^{(2)}_1 (\nn \cdot \uu^1) +\psi^{(2)}_2( \nn \cdot \uu^1) +\psi^{(2)}_{12} (\uu^1 \cdot \uu^2)  + \dots] \; \mathbf{1}. \label{explast}
\end{eqnarray}
In the present work, we will only consider the leading order terms of these expansions, and will consider the corrections in a later paper. For clarity, we can specify the functions $m_0$, $\gamma_0$ and $w_0$ in the Oseen limit, which corresponds to the situation where the two subunits are spherical with respective radii $a_1$ and $a_2$, in the limit of $x \gg a$, and neglecting the terms involving the orientation $\nn$:
\begin{eqnarray}
m_0 & \simeq & \frac{1}{6\pi\eta } \left( \frac{1}{a_1} + \frac{1}{a_2} \right) + \frac{1}{4\pi\eta x} \\
\gamma_0 & \simeq & \frac{1}{6\pi\eta } \left( \frac{1}{a_2} - \frac{1}{a_1} \right) \\
w_0 &\simeq & \frac{1}{6\pi\eta } \left( \frac{1}{a_1} + \frac{1}{a_2} \right) - \frac{1}{4\pi\eta x} 
\end{eqnarray}

\section{Average of Eq. (3)}
Starting from Eq. (3) of the main text, we use the centre of mass and elongation coordinates, $\RR=(\xx^1+\xx^2)/2$ and $\xx=\xx^2-\xx^1$, to find the Smoluchowski equation for P
\begin{eqnarray}
\partial_t P & = & \frac{\kB T}{4} \nabla_{\RR} \cdot \mathbf{M} \cdot \nabla_{\RR} P+ \frac{1}{2}\nabla_{\RR} \cdot \boldsymbol{\Gamma} \cdot (\nabla_{\xx}U)P +\frac{\kB T}{2}( \nabla_{\RR} \cdot \boldsymbol{\Gamma} \cdot \nabla_{\xx} P +  \nabla_{\xx} \cdot \boldsymbol{\Gamma} \cdot \nabla_{\RR} P) \nonumber\\
&& + \nabla_{\xx} \cdot \mathbf{W} \cdot [\kB T\nabla_{\xx} P + (\nabla_{\xx}U)P]  + \sum_{\alpha,\beta=1,2} \rotop^\alpha\cdot \mathbf{M}_\text{RR}^{\alpha \beta} \cdot [\kB T \rotop^\beta P+(\rotop^\beta U) P]  \nonumber \\
&+& \sum_{\alpha=1,2} \left\{ \nabla_\alpha \cdot \mathbf{M}_\text{TR}^{\alpha\beta} \cdot  [  (\rotop^\beta  U) P + \kB T \rotop^\beta P ] + \rotop^\alpha \cdot \mathbf{M}_\text{RT}^{\alpha \beta}  \cdot [(\nabla_\beta U) P + \kB T \nabla_\beta P]  \right\}.
\label{FP_in_R_and_x}
\end{eqnarray}
Within the expansion used in Eqs. \eqref{exp1} to \eqref{explast}, the terms involving the tensors $\mathbf{M}_\text{TR}^{\alpha\beta}$ and  $\mathbf{M}_\text{RT}^{\alpha\beta}$   will not contribute at leading order in the moment expansion that follows. We will consider these contributions in a later publication. Using the generic form for the interaction potential $U=v_0(x) + v_{01}(x) \nn\cdot \uu^1 + v_{02}(x) \nn\cdot \uu^2  +   v_{12}(x) \uu^1 \cdot \uu^2$, upon averaging Eq. (\ref{FP_in_R_and_x}) over $x$ we find
\begin{eqnarray}
\partial_t \mathcal{P} & = & \frac{\kB T}{4} \nabla_{\RR} \cdot \moy{\mathbf{M}}\cdot  \nabla_{\RR} \mathcal{P} + \frac{1}{2} \nabla_{\RR} \cdot \Bigg[\bigg<\frac{\boldsymbol{\Gamma} \, v_{01}(x)}{x}\bigg> \cdot (\mathbf{1}-\nn\nn)\cdot \uu^1 \mathcal{P} + \bigg<\frac{\boldsymbol{\Gamma} \, v_{02}(x)}{x}\bigg> \cdot (\mathbf{1}-\nn\nn) \cdot \uu^2 \mathcal{P}\Bigg] \nonumber \\
&& - \frac{\kB T}{2}\partial_{R_{i}}\ep_{jkl}\left[\mathcal{R}_l\left(\moy{\frac{\Gamma_{ij}}{x}}\hat{n}_k\mathcal{P}\right) - \moy{\mathcal{R}_l\left(\frac{\Gamma_{ij}}{x}\hat{n}_k\right)}\mathcal{P}\right] - \frac{\kB T}{2}(\nn \times \mathcal{R})\cdot(\mathbf{1}-\nn\nn)\cdot\moy{\frac{\Gamma}{x}}\cdot\nabla_{\RR}\mathcal{P}\nonumber \\
&& +\kB T(\nn \times \mathcal{R})_i(\mathbf{1}-\nn\nn)_{ij}\ep_{klm}\left[\mathcal{R}_m\left(\moy{\frac{W_{jk}}{x^2}}\hat{n}_l\mathcal{P}\right)-\moy{\mathcal{R}_m\left(\frac{W_{jk}}{x^2}\hat{n}_l\right)}\mathcal{P}\right]\nonumber\\
&& -(\nn \times \mathcal{R})\cdot(\mathbf{1}-\nn\nn)\cdot\sum_{\alpha=1,2}\moy{\frac{\mathbf{W}}{x^2}v_{0\alpha}(x)}\cdot(\mathbf{1}-\nn\nn)\cdot\uu^\alpha\mathcal{P}\nonumber \\
&& + \boldsymbol{\mathcal{R}}^1 \cdot[\langle\boldsymbol{\Psi}^{(1)} v_{12} \rangle\cdot (\uu^1 \times \uu^2)\mathcal{P} + \langle\boldsymbol{\Psi}^{(1)} v_{01} \rangle\cdot (\uu^1 \times \nn)\mathcal{P}  - \langle \mathbf{M}^{12}_\text{RR} v_{12} \rangle\cdot (\uu^1 \times \uu^2)\mathcal{P} + \langle \mathbf{M}^{12}_\text{RR} v_{02} \rangle\cdot (\uu^2 \times \nn)\mathcal{P}]\nonumber \\
&& + \boldsymbol{\mathcal{R}}^2 \cdot[-\langle\boldsymbol{\Psi}^{(2)} v_{12} \rangle\cdot (\uu^1 \times \uu^2)\mathcal{P} + \langle\boldsymbol{\Psi}^{(2)} v_{02} \rangle\cdot (\uu^2 \times \nn)\mathcal{P}  + \langle \mathbf{M}^{21}_\text{RR} v_{12} \rangle\cdot (\uu^1 \times \uu^2)\mathcal{P} + \langle \mathbf{M}^{21}_\text{RR} v_{01} \rangle\cdot (\uu^1\times \nn)\mathcal{P}],\nonumber\\
\label{LTPav_SUPP}
\end{eqnarray}
where $\mathcal{P} = \int \dd x\,  x^2 P$ and $\moy{\cdot} = \frac{1}{\mathcal{P}} \int \dd x \, x^2 \cdot P$.
We define the lowest order moments of the average distribution with respect to the three unit vectors $\nn$, $\uu^1$ and $\uu^2$ as $\rho \equiv \int_{\nn,\uu^1,\uu^2}\mathcal{P}$, $\pp \equiv \int_{\nn,\uu^1,\uu^2}\nn\mathcal{P}$ and $\pp^\alpha \equiv \int_{\nn,\uu^1,\uu^2}\uu^\alpha\mathcal{P}$ and obtain the respective evolution equations in the moment expansion of (\ref{LTPav_SUPP})
\begin{eqnarray}
\partial_t \rho & = & \frac{\kB T}{4} \moy{m_0}  \nabla^2_{\RR} \rho + \kB T \moy{\frac{\gamma_0}{x}} \nabla_{\RR} \cdot \pp + \frac{1}{3}\bigg[\bigg<\frac{\gamma_0 v_{01}}{x}\bigg>\nabla_{\RR} \cdot \pp^1 + \bigg<\frac{\gamma_0 v_{02}}{x}\bigg>\nabla_{\RR} \cdot \pp^2  \bigg],\label{momexp_rho}\\
\partial_t p_i & = & -\frac{\kB T}{3} \moy{\frac{\gamma_0}{x}} \partial_{R_i}  \rho -  2 \kB T \moy{\frac{w_0}{x^2}} p_i -\frac{2}{3}  \sum_{\alpha=1,2} \bigg<\frac{w_0 v_{0\alpha}}{x^2}\bigg> p^\alpha_{i},\label{momexp_p}\\ 
\partial_t p^\alpha_{i} & = & - 2\kB T\langle \psi_0^{(\alpha)} \rangle  p_i^{\alpha} + \frac{1}{9}\bigg<\frac{\gamma_0 v_{0\alpha}}{x}\bigg>   \partial_{R_{i}} \rho -\frac{2}{3} [\langle\psi_{0}^{(\alpha)} v_{12}\rangle p^\beta_{i} + \langle\psi_{0}^{(\alpha)} v_{0\alpha}\rangle p_{i} - \langle M_{\text{RR}\,\,0}^{\alpha\beta} v_{12}\rangle p^\beta_{i}] \label{momexp_palpha}.
\end{eqnarray}
where we have used the relation 
\begin{equation}
\moy{U' \phi(x)} = \kB T \moy{\phi'(x)+2\, \frac{\phi(x)}{x}},
\label{equilibrium_condtion}
\end{equation}
valid for any function $\phi$ of the radial coordinate and under the assumption that the $x$-dependence of $\mathcal{P}$ is Boltzmann-like i.e. $\mathcal{P} \propto \ex{-U/\kB T}$. The resulting hierarchy of equations is closed with the following prescription for the second moments
\begin{equation}
\int_{\nn,\uu^1,\uu^2}\mathcal{P}n_i n_j \simeq \frac{1}{3} \delta_{ij}; \,\,\,\,\,\, \int_{\nn,\uu^1,\uu^2}\mathcal{P}u^\alpha_i u^\beta_j \simeq \frac{1}{3} \delta^{\alpha\beta}\delta_{ij}; \,\,\,\,\,\, \int_{\nn,\uu^1,\uu^2}\mathcal{P}u^\alpha_i n_j \simeq 0,
\end{equation}
and similarly for higher order moments. We neglected any spatial derivatives of the polarisation fields in Eqs. \eqref{momexp_p} and \eqref{momexp_palpha}. A closed equation for the density $\rho$ can be obtained by taking the stationary limits of these equations, and we get $\partial_t\rho(\RR;t) = D_\text{eff} \nabla_{\RR}^2 \rho$. Using the specific form of the potential $U = \frac{1}{2} k (x-a)^2[1  + \sum_{\alpha=1,2} V_\alpha\; \nn\cdot \uu^\alpha + V_{12}\; \uu^1\cdot \uu^2]$, we  find the expression of $D_\text{eff}$  given in the main text [Eq. (4)].

\section{Definition of $\mathcal{K}_\alpha$}
\begin{equation}
\mathcal{K}_\alpha = \frac{1}{9}\left\{
\frac{\moys{\gamma_0 \varphi/x}^2}{\moys{\gamma_0 /x}^2} \frac{\moys{w_0/x^2}}{\moys{\psi_0^{(\alpha)}}} 
+\frac{\moys{\gamma_0 \varphi/x}}{\moys{\gamma_0 /x}} \frac{\moys{\psi_0^{(\alpha)} \varphi}}{\moys{\psi_0^{(\alpha)}}} 
-  \frac{\moys{\gamma_0 \varphi/x}}{\moys{\gamma_0 /x}} \frac{\moys{w_0 \varphi/x^2}}{\moys{\psi_0^{(\alpha)}}} 
- \frac{\moys{w_0\varphi/x^2}}{\moys{w_0 /x^2}} \frac{\moys{\psi_0^{(\alpha)}\varphi}}{\moys{\psi_0^{(\alpha)}}}\right\}
\label{Kalpha}
\end{equation}
where we define $\varphi (x)  =  \frac{1}{2} [\frac{x}{a}-1]^2$.

\section{Relation between the pre-averaged friction tensors to the pre-averaged mobility tensors}

We start from the following relation between the pre-averaged friction tensors and the pre-averaged mobility tensors
\begin{equation}
\label{ }
\begin{pmatrix}
    \moy{z^{11}_0}  &  \moy{z^{12}_0}     \\
    \moy{z^{21}_0}  &  \moy{z^{22}_0}     
\end{pmatrix}
=
\begin{pmatrix}
    \moy{m^{11}_\text{TT,0}}  &  \moy{m^{12}_\text{TT,0}}     \\
    \moy{m^{21}_\text{TT,0}}  &  \moy{m^{22}_\text{TT,0}}     
\end{pmatrix}^{-1},
\end{equation}
and, using the definitions of $\mathbf{M}$, $\mathbf{W}$ and $\mathbf{\Gamma}$ on the one hand, and $\mathbf{Z}$, $\mathbf{Y}$, and $\boldsymbol{\zeta}$ on the other hand (see main text), we find 
\begin{equation}
\label{ }
\moy{y_0} = 4 \; \frac{\moy{m_0}}{\moy{w_0}\moy{m_0}-\moy{\gamma_0}^2}
\qquad;\qquad
\moy{z_0} = 4 \; \frac{\moy{w_0}}{\moy{w_0}\moy{m_0}-\moy{\gamma_0}^2}
\qquad;\qquad
\moy{\zeta_0} = 4 \; \frac{\moy{\Gamma}}{\moy{w_0}\moy{m_0}-\moy{\gamma_0}^2}.
\end{equation}

\section{Binding of the enzyme to substrate and product molecules}

 The simplified mechanochemical cycle we present in the text is the following:
\begin{equation}
\label{ }
\text{E + S} \xrightleftharpoons[\beta ]{\,\alpha(S)\,} \text{C}\xrightleftharpoons[\delta(P) ]{\gamma} \text{E + P},
\end{equation}
where C represents the enzyme either bound to a substrate or a product molecule. We first determined the concentrations of S and P in the equilibrium state, in which the following conditions are satisfied:
\begin{eqnarray}
\alpha(S) p_\text{f} & = & \beta(1- p_\text{f} )\\
\gamma (1-p_\text{f} ) & = & \delta (P) p_\text{f} ,
\end{eqnarray}
where $p_\text{f}$ is the probability to find the enzyme in its free state. $\alpha(S)$ increases linearly with $S$ as $\alpha(S)=\alpha_0 S$, and $\delta(P)$ increases linearly with $P$ as $\delta(P)=\delta_0 P$, so that the solution of the system gives the expression of $S/P$ at equilibrium:
\begin{equation}
\label{ }
\frac{S}{P} = \frac{\delta_0 \beta}{\alpha_0 \gamma}.
\end{equation}

If we denote by $U_\text{S}$  (resp. $U_\text{P}$) the  conformational energy of the enzyme when it is bound to a substrate (resp. product) molecule, we write the probability of a given conformation $\CC$ under the form
\begin{equation}
p(\CC) \propto \ex{-U_\text{f}/\kB T} + \frac{S}{K_{\text{S},0}}\ex{-U_\text{S}/\kB T} + \frac{P}{K_{\text{P},0}}\ex{-U_\text{P}/\kB T} ,
\end{equation}
where we defined $K_{\text{S},0} = \beta /\alpha_0$ and $K_{\text{P},0} = \gamma /\delta_0$. For any function of a conformational state $\Phi(\CC)$, its average writes
\begin{equation}
\moy{\Phi} = \frac{\int_{\CC} \Phi\, \ex{-U_\text{f}/\kB T} + \frac{S}{K_{\text{S},0}} \int_{\CC} \Phi\, \ex{-U_\text{S}/\kB T} + \frac{P}{K_{\text{P},0}} \int_{\CC} \Phi\, \ex{-U_\text{P}/\kB T}  }{Z_\text{f} + \frac{S}{K_{\text{S},0}} Z_\text{S} + \frac{P}{K_{\text{P},0}} Z_\text{P} },
\end{equation}
with $Z_\text{X} = \int_{\CC}\ex{-\beta U_\text{X}}$. We then deduce 
\begin{equation}
\moy{\Phi} = \moy{\Phi}_\text{f} + \frac{S}{S+\frac{K_\text{S}}{K_\text{P}}P + K_\text{S}} \left[\moy{\Phi}_\text{S} -\moy{\Phi}_\text{f} \right]  + \frac{P}{P+\frac{K_\text{P}}{K_\text{S}}S + K_\text{P}} \left[\moy{\Phi}_\text{P} -\moy{\Phi}_\text{f} \right],
\end{equation}
where we define 
\begin{equation}
K_\text{S} = K_{\text{S},0} \frac{\int_{\CC} \ex{-\beta U_\text{f}}}{\int_{\CC} \ex{-\beta U_\text{S}}} \qquad\qquad;\qquad\qquad   K_\text{P} = K_{\text{P},0} \frac{\int_{\CC} \ex{-\beta U_\text{f}}}{\int_{\CC} \ex{-\beta U_\text{P}}}.
\end{equation}
This result can be written under a simpler form if we assume that $U_\text{P} = U_\text{S} = U_\text{b}$:
\begin{equation}
\label{boundfree_weight}
\moy{\Phi} = \moy{\Phi}_\text{f} + \underbrace{\left(  \frac{S}{S+\frac{K_\text{S}}{K_\text{P}}P + K_\text{S}}+ \frac{P}{P+\frac{K_\text{P}}{K_\text{S}}S + K_\text{P}} \right)}_{\equiv f(S,P)} \left[\moy{\Phi}_\text{b} -\moy{\Phi}_\text{f} \right],
\end{equation}
which defines the function $f(S,P)$ introduced in the main text. We finally consider two limits of this result:
\begin{itemize}
	\item at short times, very few product molecules have been formed ($P\simeq 0$, $S\simeq S_0$), and we find
	\begin{equation}
	\moy{\Phi} = \moy{\Phi}_\text{f} +  \frac{S_0}{S_0+K_\text{S}}  \left[\moy{\Phi}_\text{b} -\moy{\Phi}_\text{f} \right],
	\end{equation}
	which corresponds to the expression given in the main text, where we neglected the formation of product molecules.
	\item at long times, when chemical equilibrium is reached, the concentrations of $S$ and $P$ are given by
	\begin{equation}
	S = \frac{K_\text{S}}{K_\text{S}+K_\text{P}}S_0 \qquad\qquad;\qquad\qquad  P = \frac{K_\text{P}}{K_\text{S}+K_\text{P}}S_0,
	\end{equation}
	and we find
	\begin{equation}
	\moy{\Phi} = \moy{\Phi}_\text{f} +  \frac{S_0}{S_0+\frac{1}{2}(K_\text{S}+K_\text{P})}  \left[\moy{\Phi}_\text{b} -\moy{\Phi}_\text{f} \right],
	\label{boundfree_weight_longtimes}
	\end{equation}
\end{itemize}
so that the effective equilibrium constant used in the main text reads
\begin{equation}
\label{ }
K = \frac{1}{2}(K_\text{S}+K_\text{P}) = \frac{1}{2} \left[ K_{\text{S},0} \frac{\int_{\CC} \ex{-\beta U_\text{f}}}{\int_{\CC} \ex{-\beta U_\text{S}}} +   K_{\text{P},0} \frac{\int_{\CC} \ex{-\beta U_\text{f}}}{\int_{\CC} \ex{-\beta U_\text{P}}}  \right]   .
\end{equation}

\section{Contributions to the dimensionless coefficient $\mathcal{A}$}

\subsection{Change in the diffusion coefficient}

Here, we  estimate the change in the diffusion coefficient of the dumbbell due to the structural modifications induced by substrate binding presented above. Neglecting the effect of the orientational degrees of freedom in Eq. (4) from the main text, we write the effective diffusion coefficient of the enzyme for an arbitrary substrate concentration as
\begin{equation}
\label{Dsimple_supplement}
D(S)  = \frac{\kB T}{4}  \moy{m_0}- \frac{\kB T}{6} \frac{\moy{\gamma_0/x}^2}{\moy{w_0/x^2}},
\end{equation}
where the averages are to be understood with the weight defined in Eq. \eqref{boundfree_weight}.
The change in the diffusion coefficient is given as
\begin{equation}
\label{deltaDgeneric}
\Delta D =  D(S)-D(S=0)= \left[  \frac{\kB T}{4}  \moy{m_0}- \frac{\kB T}{6} \frac{\moy{\gamma_0/x}^2}{\moy{w_0/x^2}} \right]-\left[  \frac{\kB T}{4}  \moy{m_0}_\text{f} - \frac{\kB T}{6} \frac{\moy{\gamma_0/x}_\text{f}^2}{\moy{w_0/x^2}_\text{f}} \right],
\end{equation}
where indexed brackets denote an average with the probability distribution corresponding to the free state as defined above.
Using Eq. \eqref{boundfree_weight_longtimes} to write the average quantities in terms of the corresponding pure averages in the free and bound states in the long-time limit, we find
\begin{equation}
\label{deltaDgeneric}
\frac{\Delta D}{D_0}=\mathcal{A} \cdot \frac{S_0}{S_0+K}, 
\end{equation}
where
\begin{equation}
\label{AdeltaDgeneric}
\mathcal{A}=\frac{\left[ \frac{1}{4}  \moy{m_0}_\text{f} \left(\frac{ \moy{m_0}_\text{b}- \moy{m_0}_\text{f}}{ \moy{m_0}_\text{f}}\right)\right]- \frac{1}{6} \frac{\moy{\gamma_0/x}_\text{f}^2}{\moy{w_0/x^2}_\text{f}}\left[2 \left(\frac{\moy{\gamma_0/x}_\text{b}-\moy{\gamma_0/x}_\text{f}}{\moy{\gamma_0/x}_\text{f}}\right)-\left(\frac{\moy{w_0/x^2}_\text{b}-\moy{w_0/x^2}_\text{f}}{\moy{w_0/x^2}_\text{f}}\right) \right]}{\left[\frac{1}{4}  \moy{m_0}_\text{f} - \frac{1}{6} \frac{\moy{\gamma_0/x}_\text{f}^2}{\moy{w_0/x^2}_\text{f}}\right]}, 
\end{equation}
to the lowest order in difference between free and bound averages. We now consider the different effects separately.

\subsection{Modification of the equilibrium distance between the subunits}

Substrate binding is expected to reduce the equilibrium distance between the subunits that constitute the dumbbell. We first estimate the contribution of this effect to the diffusion coefficient of the dumbbell, by using the explicit (harmonic) form for the potentials $U_\text{f}$ and $U_\text{b}$:
\begin{equation}
\label{ }
U_\text{f} = \frac{1}{2} k_\text{f} (x-a)^2 \qquad ; \qquad U_\text{b} = \frac{1}{2} k_\text{f} [x-(a-\delta x)]^2,
\end{equation}
and expand the expression of $\Delta D/D_0$ [Eq. \eqref{deltaDgeneric}] in the small deformation limit ($\delta x \ll a$), in which  the following expression 
\begin{equation}
\label{ }
\moy{\Phi}_\text{b}-\moy{\Phi}_\text{f}=\frac{k_\text{f} \, \delta x}{\kB T} \left( \moy{x}_\text{f} \moy{\Phi}_\text{f}  - \moy{x\Phi}_\text{f}  \right) + \mathcal{O}[(\delta x)^2] ,
\end{equation}
is to be incorporated for each average quantity. 
At linear order in $\delta x$, we find the change in the diffusion coefficient of the enzyme due to its compression
\begin{equation}
\label{DD2}
\mathcal{A}_\text{c}\simeq\frac{\left[ \frac{1}{4}  \moy{m_0}_\text{f} \left(\moy{x}_\text{f} \moy{m_0}_\text{f}  - \moy{x m_0}_\text{f}\right)\right]- \frac{1}{6} \frac{\moy{\gamma_0/x}_\text{f}^2}{\moy{w_0/x^2}_\text{f}}\left[2 \left(\moy{x}_\text{f} \moy{\gamma_0/x}_\text{f}  - \moy{\gamma_0}_\text{f} \right)-\left(\moy{x}_\text{f} \moy{w_0/x^2}_\text{f}  - \moy{w_0/x}_\text{f} \right) \right]}{\left[\frac{1}{4}  \moy{m_0}_\text{f} - \frac{1}{6} \frac{\moy{\gamma_0/x}_\text{f}^2}{\moy{w_0/x^2}_\text{f}}\right]}\; \frac{k_\text{f} \, \delta x}{\kB T}. 
\end{equation}
In the limit where the stiffness of the spring linking the two subunits of the enzyme is very large ($k_\text{f} \gg \kB T/a^2$), we can expand the averages $\moy{\cdot}_\text{f}$ using Laplace's method and rewrite the amplitude of the relative increase of the diffusion coefficient as
\begin{equation}
\label{DeltaDfinal}
\mathcal{A}_\text{c} =  \mathcal{G} \cdot \frac{\delta x}{a},
\end{equation}
where
\begin{equation}
\mathcal{G} = a \left[ -m_0' +\frac{1}{6} \frac{\gamma_0^2}{w_0}  \left( \ln \frac{\gamma_0^2}{w_0} \right)'\right] \left[m_0 - \frac{1}{6}\frac{\gamma_0^2}{w_0}  \right]^{-1},
\label{Gdef}
\end{equation}
and the mobility coefficients and their derivatives are evaluated at the equilibrium position $x=a$.

\subsection{Modification of the potential stiffness}

Using the explicit expressions of the potentials $U_\text{f}$ and $U_\text{b}$ 
\begin{equation}
\label{ }
U_\text{f} = \frac{1}{2} k_\text{f} (x-a)^2 \qquad ; \qquad U_\text{b} = \frac{1}{2} k_\text{b} (x-a)^2,
\end{equation}
in Eq. \eqref{deltaDgeneric}, we can calculate the contribution due to change in stiffness.

This contribution can first be estimated numerically, by using different approximations for the mobility coefficients of the subunits of the dumbbell, when they are assumed to be spherical. In the far-field (Oseen) limit, these mobility coefficients are given by \cite{Kim2005}
\begin{equation}
\label{ }
\mathbf{M}^{\alpha\alpha}_\text{TT} = \frac{1}{6\pi\eta a_\alpha} \mathbf{1} 
\qquad;\qquad
\mathbf{M}^{\alpha\beta}_\text{TT} = \frac{1}{8\pi\eta x} ( \mathbf{1} + \nn \nn) \qquad (\alpha\neq\beta),
\end{equation}
where $a_\alpha$ is the radius of subunit $\alpha$. Defining $\mathbf{M}_\text{TT}^{\alpha\beta} = {M}^{\alpha\beta}_{\text{I}} \mathbf{1}+{M}^{\alpha\beta}_{\text{D}} \nn\nn$, the corrections to these leading order terms can be obtained as a series expansion in powers of $1/x$ \cite{Jeffrey1984}:
\begin{eqnarray}
{M}^{\alpha\alpha}_{\text{I}}  & = & \frac{1}{6\pi \eta a_\alpha} - \frac{17}{96}\frac{a_\beta^5}{\pi\eta x^6} - \frac{1}{48}\frac{a_\beta^3(10 a_\alpha^4-9a_\alpha^2 a_\beta^ 2 + 9 a_\beta^4)}{\pi\eta x^8}  
- \frac{1}{32}\frac{a_\beta^5(35 a_\alpha^4-18 a_\alpha^2 a_\beta^2 + 6 a_\beta^4)}{\pi\eta x^{10}},   \\
{M}^{\alpha\alpha}_{\text{D}}  & = & -\frac{5}{8} \frac{a_\beta^3}{\pi \eta x^4} + \frac{5}{32} \frac{(8 a_\alpha^2-a_\beta^2)a_\beta^3}{\pi\eta x^6}-\frac{1}{48} \frac{(20 a_\alpha^4-123 a_\alpha^2a_\beta^2+9a_\beta^4)a_\beta^3}{\pi\eta x^8}\nonumber\\
&&-\frac{1}{96} \frac{(175 a_\alpha^4 +1500 a_\alpha^3a_\beta-426 a_\alpha^2a_\beta^2+18 a_\beta^4)a_\beta^5}{\pi\eta x^{10}},\\
{M}^{\alpha\beta}_{\text{I}}  & = & \frac{1}{8 \pi \eta x} + \frac{1}{24} \frac{a_\alpha^2+a_\beta^2}{\pi\eta x^3} + \frac{7}{768}\frac{(80a_\alpha^4-79a_{\alpha}^2a_{\beta}^2+80a_{\beta}^4)a_{\alpha}^2a_\beta^3}{\pi\eta x^{11}},\\
{M}^{\alpha\beta}_{\text{D}}  & = & \frac{1}{8 \pi \eta x} -\frac{1}{8} \frac{a_\alpha^2+a_\beta^2}{\pi\eta x^3} + \frac{25}{8}\frac{a_\alpha^3 a_\beta^3}{\pi\eta x^7} - \frac{5}{8}\frac{(a_\alpha^2+a_\beta^2)a_\alpha^3a_\beta^3}{\pi\eta x^9} + \frac{1}{768}\frac{(400a_\alpha^4-13943a_\alpha^2a_\beta^2+400a_\beta^4)a_\alpha^3 a_\beta^3}{\pi\eta x^{11}}.
\end{eqnarray}
In the Smoluchowski description of the stochastic dynamics of the system, and  for the particular case of spherical subunits, we can refine the approximate forms of the mobility tensors and use $\mathbf{A} = A_\text{I} \mathbf{1} + A_\text{D} \nn \nn$ instead of $\mathbf{A} \simeq  a_0 \mathbf{1}$ for $\mathbf{A}=\mathbf{M},\mathbf{W}$ and $\boldsymbol{\Gamma}$. We will consider this approximation in greater details in a later publication.  We can show that Eq. \eqref{Dsimple_supplement} still holds, with $m_0 = M_\text{I}+M_\text{D}/3$, $\gamma_0=\Gamma_\text{I}$ and $w_0=W_\text{I}$. These expressions for the mobility functions are used to produce the plots shown in the main text.

We can also estimate analytically the relative change of the diffusion coefficient due to an increase in the potential stiffness in the limit of $k_\text{f}\to\infty $ and $k_\text{b}\to\infty $ with a fixed difference $\delta  k  = k_\text{b} - k_\text{f}$:
\begin{equation}
\label{DeltaDfinal}
\mathcal{A}_\text{s}=  \mathcal{H} \cdot \varepsilon^2 \cdot \frac{\delta k}{k_\text{f}},
\end{equation}
where $\varepsilon = \sqrt{\kB T/(k_\text{f} a^2)}$ is a dimensionless number that   characterises the amplitude of the thermal fluctuations of the dumbbell elongation around its equilibrium value, and where the dimensionless coefficient $\mathcal{H}$ reads
\begin{equation}
\label{ }
\mathcal{H} = - \frac{1}{2} \left[ a^2\frac{m_0''}{4}  + am_0' + \frac{1}{3}\frac{\gamma_0^2}{w_0} -\frac{2}{3}a \frac{\gamma_0\gamma_0'}{w_0} - \frac{1}{3} a^2 \frac{\gamma_0 \gamma_0''}{w_0} + a^2 \frac{w_0'' \gamma_0^2}{6 w_0^2}\right]\left[\frac{m_0}{4}-\frac{1}{6}\frac{\gamma_0^2}{w_0}\right]^{-1},
\end{equation}
with the mobility coefficients and  derivatives  evaluated at the equilibrium position $x=a$.

\subsection{Hindering of orientational fluctuations}

We finally estimate the contribution to $\mathcal{A}$ coming from changes in orientational fluctuations through the dimensionless coefficients $V_\alpha$. Denoting by $V_{\alpha,\text{f}}$ (resp. $V_{\alpha,\text{b}}$) the value of the coefficients $V_\alpha$ in the free (resp. bound) state, we find at leading order in the corrections $[\moy{\Phi}_\text{b}-\moy{\Phi}_\text{f}]$ the contribution to the dimensionless coefficient $\mathcal{A}$ that originates from constraints on the orientational fluctuations of the subunits:
\begin{equation}
\label{ }
\mathcal{A}_\text{r}     = \frac{ \frac{\kB T}{3} \frac{\moy{\gamma_0/x}_\text{f}^2}{\moy{w_0/x^2}_\text{f}} \sum_{\alpha=1,2}   \left( \frac{k_\text{f} a^2}{\kB T} \right)^2 V_{\alpha,\text{f}}(V_{\alpha,\text{b}}-V_{\alpha,\text{f}})\mathcal{K}_\alpha  }{ \frac{\kB T}{4}  \moy{m_0}_\text{f}- \frac{\kB T}{6} \frac{\moy{\gamma_0/x}_\text{f}^2}{\moy{w_0/x^2}_\text{f}} \left[ 1-\sum_{\alpha=1,2} \left( \frac{k_\text{f} a^2}{\kB T} V_\alpha \right)^2 \mathcal{K}_\alpha \right] }
\end{equation}
As for the coefficients $\mathcal{A}_\text{c}$ and $\mathcal{A}_\text{s}$, this expression can be estimated in the limit where $k_\text{f} \gg \kB T/a^2$:
\begin{equation}
\label{ }
\mathcal{A}_\text{r} \simeq \frac{\kB T }{k_\text{f} a^2} \sum_{\alpha=1,2} V_{\alpha,\text{f}} (V_{\alpha,\text{b}}-V_{\alpha,\text{f}}) \mathcal{J}_\alpha
\end{equation}
with
\begin{equation}
\label{ }
\mathcal{J}_\alpha = \frac{\frac{1}{18}  \frac{\gamma_0^2}{w_0}  }{\frac{m_0}{4}-\frac{1}{6} \frac{\gamma_0^2}{w_0}}   \frac{(a^2 \psi_0^{(\alpha)}+w_0)  (\gamma_0'' w_0 a -\gamma_0 w_0'' a + 2 \gamma_0' w_0)    }{a   \gamma_0 w_0 \psi_0^{(\alpha)}}  
\end{equation}
with the mobility coefficients and  derivatives  evaluated at the equilibrium position $x=a$.

\end{document}